\documentclass[%
 reprint,
 amsmath,amssymb,
 aps,
prb,
]{revtex4-2}
\usepackage[backref=none]{hyperref}
\usepackage{graphicx}
\usepackage{dcolumn}
\usepackage{comment}
\usepackage{lipsum}
\usepackage{xcolor}
\usepackage{physics}
\usepackage{bm}
\usepackage{tikz}
\usetikzlibrary{trees}
\usepackage{soul}
\begin{document}

\preprint{APS/123-QED}

\title{Multimode NOON-state generation with ultracold atoms via geodesic counterdiabatic driving}
\author{Simon Dengis$^{1}$}
\author{Sandro Wimberger$^{2,3}$}
\author{Peter Schlagheck$^{1}$}
\affiliation{
  $^{1}$CESAM research unit, University of Liege, B-4000 Li\`ege, Belgium}
\affiliation{
 $^{2}$Dipartimento di Scienze Matematiche, Fisiche e Informatiche,
  Universit\`a di Parma, Parco Area delle Scienze 7/A, 43124 Parma, Italy}
\affiliation{
$^{3}$INFN, Sezione di Milano Bicocca, Gruppo Collegato di Parma,
Parco Area delle Scienze 7/A, 43124 Parma, Italy
}%

\begin{abstract}

We present a protocol for the generation of NOON states with ultracold atoms, leveraging the Bose-Hubbard model in the self-trapping regime. By the means of an optimized adiabatic protocol, we achieve a significant reduction in the time required for the preparation of highly entangled NOON states, involving two or more modes. Our method saturates the quantum speed limit, ensuring both efficiency and high fidelity in state preparation. A detailed analysis of the geodesic counterdiabatic driving protocol and its application to the Bose-Hubbard system highlights its ability to expand the energy gap, facilitating faster adiabatic evolution. Through perturbation theory, we derive effective parameters that emulate the counterdiabatic Hamiltonian, enabling experimentally viable implementations with constant physical parameters. This approach is demonstrated to yield exponential time savings compared to standard geodesic driving, making it a powerful tool for creating complex entangled states for applications in quantum metrology and quantum information. Our findings pave the way for scalable and precise quantum state control in ultracold atomic systems.
\end{abstract}

\maketitle

\section{Introduction}
Quantum entanglement \cite{EPR} is arguably one of the phenomena that has garnered the most attention and debate among physicists over the last couple of decades. Since the end of the 20th century, this unique feature of quantum mechanics has been the focal point of endless discussions among some of the greatest minds in modern science. Today, entanglement plays a critical role in the field of quantum information \cite{NieChu}. Specifically, the realization of entangled states with a large number of particles could significantly enhance the robustness of these states against external disturbances such as decoherence and noise. Among these entangled states, NOON states, which consist of a coherent superposition of states that have $N$ particles in two orthogonal modes, $\vert N,0\rangle $ and $\vert 0,N\rangle$, are of particular interest for quantum metrology and sensing \cite{sensing1,sensing4, metrologie, sensing3, sensing2,Panda2024}. To date, NOON states have been realized using photons and phonons for particle numbers $N\leq 10$ \cite{LightNOON,SCQB,Phonons}.
\begin{figure}[!t]
    \centering
    \includegraphics[width=0.475\textwidth]{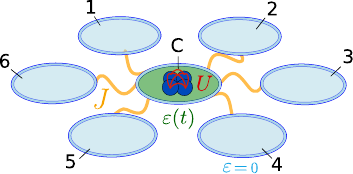}
    \caption{Schematic representation of the star shaped Bose-Hubbard model for $L=6$  and a central initially populated well. The bosons can tunnel from the central well to any of the outer wells with a hopping rate $J$. Particles localized in the same well present an interaction $U$. Initially, all bosons are localized in the central well, whose energy $\varepsilon$ is isolated far from the rest of the spectrum. Over time, the energy of the central well is varied by a function $\varepsilon (t)$ to induce an adiabatic transfer of the system’s state to the desired coherent superposition $\vert \text{L-NOON}\rangle$. At the end of the protocol, the central well is again isolated from the rest of the spectrum, leaving only the entangled state across all $L$ wells.}
    \label{stargraph}
\end{figure}

The creation of such entangled states within the framework of ultracold atoms has been intensively discussed \cite{Bec1,Bec3,Bec2,Bec4,Bec5,Bec6,Fischer2015,Bycheck2018,Pezze2019,protocolNOON,Coll3,Gui1,Gui2,Beringer2024,recentNOON}. A notable feature of atomic gases is the interaction between particles, which allows certain states to be isolated within the energy spectrum of the system. For instance, when particles interact, states where all particles are localized in the same mode tend to cluster together, separated from the rest of the spectrum. This suppresses sequential tunneling and thus gives rise to the phenomenon of collective tunneling \cite{Coll1,Coll2,Coll4,Coll3}, where particles undergo tunneling between different modes collectively. This behavior has enabled the first creation of NOON like states in systems of atoms trapped in a perfectly symmetric double-well potential \cite{recentNOON}, albeit with a rather limited purity. 

In the deep self-trapping regime \cite{selftrap}, where particle interactions significantly exceed the hopping rate between sites, a NOON state is formed after half the time required for complete tunneling. However, this method poses a challenge: since the tunneling time is inversely proportional to the energy difference between the two modes, the time required to observe the creation of a NOON state far exceeds the lifetime of a Bose-Einstein condensate. To address this limitation, a faster process was proposed, leveraging chaos-assisted and resonance-assisted tunneling \cite{Tomsovic1994,Brodier2001,peter1}, which reduced the time required for NOON states creation by several orders of magnitude \cite{Gui1}.  This same method has been applied to create triple-NOON states in three-site systems \cite{Gui2}, but the time needed to entangle the system is still long ($\sim 1$s for ultracold $^{87}$Rb gases with $N = 5$).

One possible method to enhance the tunneling, and thus to overcome the time scale problem, involves the adiabatic theorem \cite{Chen,Campo,Deffner2014,delCampo2019,ieva,Dengis25}. This theorem states that if a system is perturbed very slowly and there exists an energy gap between its eigenvalues and the rest of the spectrum, the system’s state will follow the eigenstates of its Hamiltonian over time. Consider a system of ultracold atoms in $L$ modes, along with a central mode symmetrically coupled to the 
$L$ others (see Fig.~\ref{stargraph}) \cite{starBH}. If all atoms are initially placed in the central well, situated at the top of the system’s energy spectrum, a sufficiently slow driving toward a lower energy state corresponding to the $L$ other modes could result in the creation of a $L$-NOON state after a time $T$. However, by essence, this method has to employ a slow driving in order to steer the system to the desired state. 

The protocol speed can be enhanced by different methods, known as shortcuts to adiabaticity \cite{Chen2010,Martinez2014,Odelli2023,Odelli2024,STA}. We focus here on Counterdiabatic driving (CD) \cite{Opatrný_2014,Chen}, which has proven its effectiveness through many applications \cite{Malossi2013,Du2016,Zhou2017,Hu2018,Santos2020,Dogra2022,Bosch2025}. Moreover, if the driving is chosen such that the time evolution follows one of the geodesics of the parameter space \cite{demiplak,demirplak2,Class, geodesic}, some elements of the CD can be made time-independent, which allows for the definition of an efficient experimental protocol to follow adiabatic dynamics. This method was developed for $L=2$ in \cite{Dengis25}, where we proposed an experimental protocol for the accelerated realization of NOON states with ultracold atoms within experimentally feasible time scales, called the geodesic counterdiabatic driving (GCD).

The creation of more exotic NOON states can exhibit intriguing properties, particularly in the field of quantum metrology \cite{PRLNOON}. In this article, we generalize the method to the generation of multi-mode NOON states, with a specific focus on the accelerated creation of triple-NOON states. Furthermore, we demonstrate that the GCD method is able to reach the quantum speed limit.

The paper is structured as follows. Section II outlines the context of the studied system, namely the Bose-Hubbard model describing a set of bosonic particles trapped in a lattice. Section III details shortcut-to-adiabaticity methods, namely geodesic and counterdiabatic drivings, and unifies them in the GCD protocol. Section IV focuses on applying the driving process to several multi-mode NOON states, and on understanding its behavior with $L$. Conclusions are found in section V.
\section{Bose-Hubbard model in the self-trapping regime} 

Consider a Bose-Hubbard model (BHM) \cite{BHModel} that describes a system of $N$ bosons trapped in a lattice of L+1 sites (see Fig.~\ref{stargraph}). The bosons on the same site interact with strength $U$ and can tunnel between neighboring sites at a hopping rate $J$. In the present study, we consider a set of L sites, all coupled to a central well denoted by the letter $c$. The energy of the central well can be modulated via a time-dependent bias field with strength $\varepsilon$. Using creation ($\hat{a}^{\dagger}_{i}$) and annihilation ($\hat{a}_{i}$) operators, the Hamiltonian describing the system is expressed as:
\begin{align}\label{BHM}
    \hat{H}(t)= &\frac{U}{2}\sum_{i=1}^{L}\hat{a}_{i}^{\dagger}\hat{a}_{i}^{\dagger}\hat{a}_{i}\hat{a}_{i} 
    -J\sum_{i=1}^{L}\left(\hat{a}^{\dagger}_{c}\hat{a}_{i} + \hat{a}^{\dagger}_{i}\hat{a}_{c}\right) \\
    \notag
    &+ \frac{U}{2}\hat{a}_{c}^{\dagger}\hat{a}_{c}^{\dagger}\hat{a}_{c}\hat{a}_{c} +\varepsilon(t)\hat{a}_{c}^{\dagger}\hat{a}_{c}
\end{align}
Several regimes emerge depending on the parameters $U$ and $J$, describing phenomena such as a Mott insulator or a superfluid. The parameter region of interest is the self-trapping regime, which describes a system of strong interaction effects and weak hopping, that is, $NU/J \gg 1$ \cite{selftrap}.
\begin{figure}[!t]
    \centering
    \includegraphics[width=0.475\textwidth]{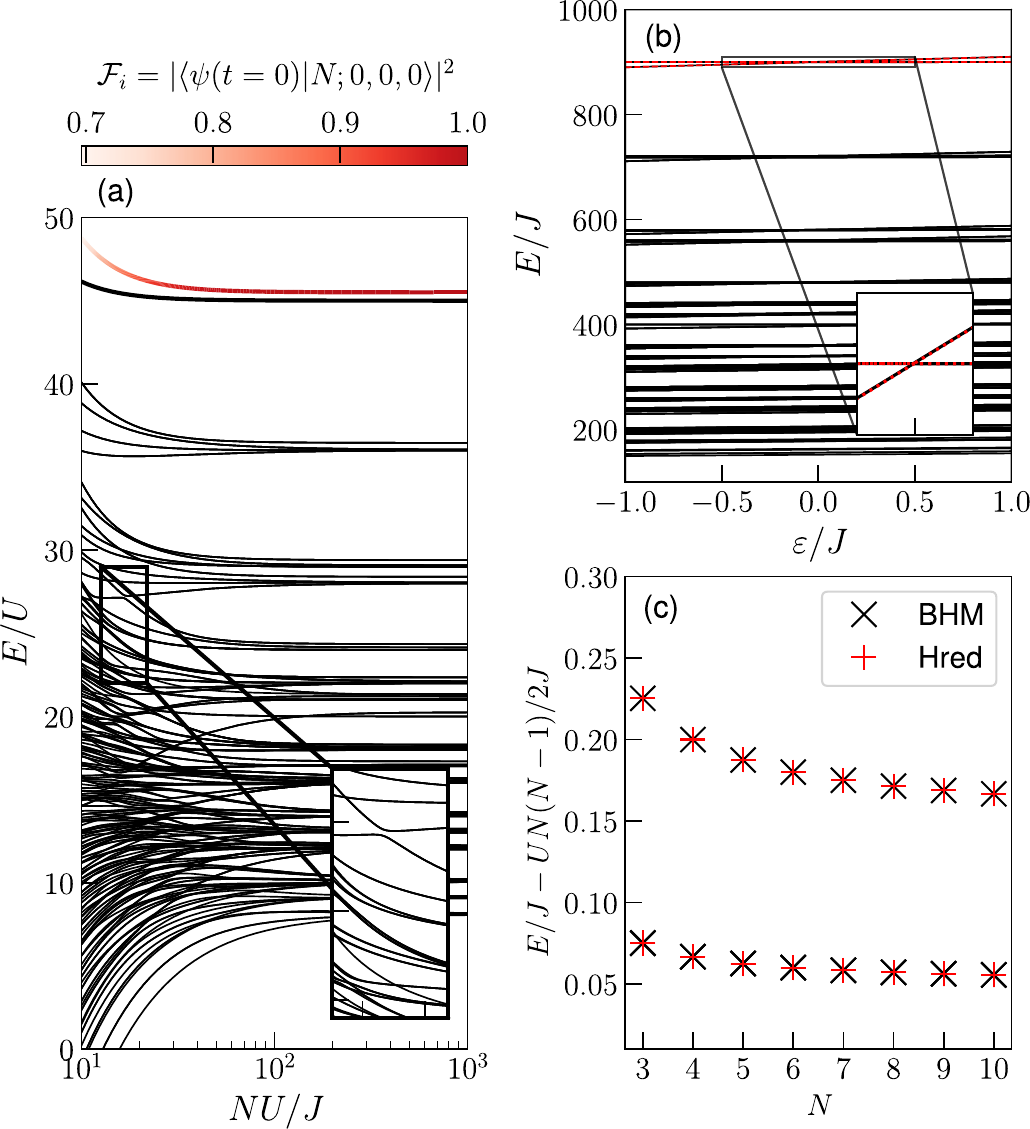}
    \caption{ 
    (a) Spectrum as a function of $NU/J$ for $N=10$, $L=3$ and $\varepsilon=0$. It is clearly possible to define a self-trapping regime, within which several effectively separable states in the energy spectrum can be identified. In these subsystems, the size of the Hilbert space is significantly reduced, allowing for the determination of a simplified Hamiltonian. The validity of the approximation can be understood as the overlap $F_{i} = \vert \langle \psi(t=0) \vert N;0,0,0\rangle\vert^{2}$, depicted in a red fade, between the initial state $\vert \psi(0)\rangle$ and the Fock state having all the bosons located in the central well. (b) Spectrum as a function of driving parameter $\varepsilon/J$ , with $N=10$, $L=3$ and $U=20J$. Once the self-trapping regime is set, the top energy levels are isolated and a reduced system can be defined. The validity of the approximation is verified in (c), where the eigenenergies of the reduced Hamiltonian are compared with the top energies of the full Bose-Hubbard Hamiltonian for several numbers of particles $N$, $L=3$ and $ U=20J$ .}
    \label{fig:enter-label}
\end{figure}

Figure 2(a) illustrates the evolution of the energy levels as a function of the ratio $NU/J$, for L=3, $N=10$ and $U=20J$. Beyond a certain value of this ratio, the highest energy levels become completely isolated from the rest of the spectrum and we are in the self-trapping regime. A gap of width $U(N-1)/J$ protects the system's highest energy levels from the influence of other states, allowing its temporal evolution to be restricted to the states nearest to the initial condition. In this self-trapping regime, an adiabatic driving can be applied to the central well to induce energy level crossings and create a gap that enables the generation of entangled states. Figure 2(b) illustrates the evolution of the spectrum as a function of the driving parameter $\varepsilon$, highlighting the presence of a gap of value $\simeq (J/U)^{N-1}$ for $L=4$, $N=10$, and $U=20J$.  

\subsection*{Perturbation theory}
\begin{figure}[t]
    \centering
    \includegraphics[width=0.475\textwidth]{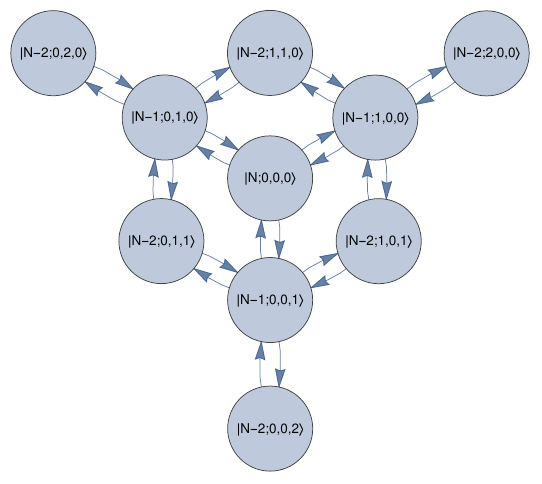}
    \caption{Perturbative diagram of the energy corrections to the state $\vert N;0,0,0\rangle$, for $L=3$, at fourth order. Each path contributes to the state's energy in the manner of a Feynman path integral. Each arrow represents a possible transition path and carries a probabilistic amplitude proportional to $J/U$. The correction to the energy of the state $\vert N;0,0,0\rangle$ is given by the sum over all paths of length $l$—where $l$ corresponds to the desired perturbative order—that start at and return to this state. The number of such paths (24 in this example) is determined by the $l^{\text{th}}$ power of the adjacency matrix associated with the diagram.}
    \label{diag}
\end{figure}

Eliminating from the dynamics those levels that are not affected by the driving, we can work within a two-level basis, consisting of the states $\vert N;0,0,...\rangle$ and $(\vert 0;N,0,...\rangle + \vert 0;0,N,...\rangle + ...)/\sqrt{L}$. The reduced Hamiltonian in this basis is expressed as:
\begin{align}\label{Hred}
    H_{\text{red}}(U,J,\varepsilon(t))= \left(
\begin{tabular}{c c}
 $\Tilde{\mathcal{E}} + \mathcal{N}\varepsilon(t)$ & $-\sqrt{L}\mathcal{J}$  \\ 
 $-\sqrt{L}\mathcal{J}$ & $\mathcal{E}$ \\
\end{tabular}
\right)
\end{align}
where $\mathcal{E}=\mathcal{E}(U,J)$, $\Tilde{\mathcal{E}} = \Tilde{\mathcal{E}}(U,J)$, $\mathcal{N} = N + \delta N(U,J)$ and $\mathcal{J} = \mathcal{J}(U,J)$ are obtained using perturbation theory for $NU/J \gg 1$. 
The method employed in the Supplemental Material of \cite{Dengis25} provides the derivation of these parameters through the perturbative solution of the time-independent Schrödinger equation. It is possible to visualize this approach by directly analyzing the probabilities associated with the particles following different paths within the Hilbert space. Assuming that all particles are initially located in the central well, we can determine the various trajectories that the state of the system may take (see Fig. \ref{diag}). For an arbitrary number of $L$ external sites, their expressions at the 4th order of perturbation theory following this approach are fully determined as:
\begin{align}
    \Tilde{\mathcal{E}}(U,J) = \frac{NLJ^{2}}{(N-1)U} -\frac{NLJ^{4}}{(N-1)^{3}(N-2)U^{3}}\\
    \notag \times \frac{(L+1)(N-2)^{2}-2N(N-3)-5}{(2N-3)},
\end{align}
\begin{align}
    &\mathcal{E}(U,J) = \frac{NJ^{2}}{(N-1)U}+ \frac{N((L+1)(N-2) - 2N +5)J^{4}}{(N-1)^{3}(N-2)U^{3}},\\
    \label{couplageeff}&\mathcal{J}(U,J) = \frac{NJ^{N}}{(N-1)!U^{N-1}}.
\end{align}
\noindent Hamiltonian (\ref{Hred}) faithfully reproduces the effective two-level dynamics of the full Bose-Hubbard Hamiltonian in the self-trapping regime. The system reduction provides valuable insights into the structure of the gap created between the states $\vert N;0,0,...\rangle$ and $\vert \text{L-NOON}\rangle \equiv (\vert 0;N,0,...\rangle + \vert 0;0,N,...\rangle + ...)/\sqrt{L}$. In particular, the evolution of the spectrum of $H_{\text{red}}$ is fully determined, and the minimum separation between the two energy levels, corresponding to the minimal gap, is $2\sqrt{L}\vert\mathcal{J}\vert$. In Fig.~2(c) the validity of the approximation is verified for $ L=3$, $U=20J$ and up to $N=10$, where the top four energies of the Bose-Hubbard Hamiltonian (\ref{BHM}) are compared with the eigenvalues of the reduced Hamiltonian (\ref{Hred}). A complementary verification is yielded in Appendix A, where the next order in perturbation theory is calculated. The goal now is to exploit this gap to induce entanglement between the states via the adiabatic theorem.
\section{Shortcut to adiabaticity}

\subsection{Geodesic driving}
The adiabatic theorem \cite{Born} states that when a system undergoes a time-dependent perturbation whose variation in amplitude is smaller than the system's intrinsic frequency scale corresponding to level splitting in the spectrum, the system will, if it initially has been prepared in an eigenstate of the Hamiltonian, follow that eigenstate in the course of time evolution. Most often, adiabatic following of a system's eigenstate $\vert n\rangle$ can be granted \cite{Landau,Zener,Stuck,Majorana,Berry1987,Unanyan1997,Fleischhauer1999,Lim1991} if
\begin{equation}
    \frac{\vert \langle m\vert \dot{H}\vert n\rangle\vert}{ (E_{n}-E_{m})^{2}} \ll \hbar^{-1},
\end{equation}
for all $m \neq n$.

If this criterion is sufficient to determine a satisfactory driving protocol to adiabatically guide a system, it is possible to determine the optimal path to minimize the protocol time \cite{Provost, geodesic,Carlini2006,Nauth2022}. To achieve this, it is necessary to identify the geodesics of the system within the parameter space. In this context, the use of differential geometry allows for the description of the distance between two states in Hilbert space as a function of a small variation in the parameter $\varepsilon$: $ds^2 = 1 - \vert\bra{n(\varepsilon)}\ket{n(\varepsilon + d\varepsilon})\vert^2 = g^{(n)} d\varepsilon^{2}$,
where $g^{(n)}$ is, here, the only non-vanishing component of the real part of the quantum geometric tensor defined in terms of the instantaneous eigenstates $\vert m\rangle$ of the Hamiltonian matrix $H(t)$ and their associated eigenvalues $E_{m}$, 
\begin{equation}\label{gmunu}
        g^{(n)} = \text{Re} \sum_{m\neq n}  \frac{\braket{n |\partial_{\varepsilon} H_{\text{red}}}{m} \braket{m | \partial_{\varepsilon} H_{\text{red}}}{n}  } {(E_{n} - E_{m})^{2}} 
\end{equation}
Rigorously, $g^{(n)}$ is the Riemann metric on the parameter manifold $\mathcal{M}$, since $\varepsilon$ is the time-dependent parameter that drives the system. The fact that $\mathcal{M}$ is a metric space provides the notion of geodesic curves, which are paths that minimize the distance functional $\mathcal{L}(\varepsilon) =$ $\int_{\varepsilon(t=t_{i})}^{\varepsilon(t=T)}ds =$
$\int_{t_{i}}^{T}\sqrt{g^{(n)}\dot{\varepsilon}^{2}}dt$
between two points $\varepsilon(t=t_{i})\equiv \varepsilon_{i}$ and $\varepsilon(t=T)\equiv \varepsilon_{f}$, with $\dot{\varepsilon}  \equiv d\varepsilon/dt$. The integrand of $\mathcal{L}$ correspond to the fidelity $\mathcal{F}$ between infinitesimally separated states. Applying the least action principle, Euler-Lagrange equations gives a geodesic equation $g^{(n)}\dot{\varepsilon}(t)^{2}=\text{const.}$, describing the path that should follow $\varepsilon(t)$ in order to minimize the infidelity \cite{demiplak, demirplak2, Class, geodesic}. The metric tensor defines then the leading non-adiabatic correction to the energy variance: $\Delta E^{2} = -\hbar^{2}g^{(n)}\dot{\varepsilon}^{2}(t)$  \cite{Kolodrubetz2017}. 

For our reduced Hamiltonian $H_{\text{red}}$, the geodesic protocol calculation is straightforward and yields: 
\begin{equation}
    \mathcal{N}\varepsilon(t) = 2\sqrt{L}\mathcal{J}\text{tan}\left( \alpha_{i} + (\alpha_{f}-\alpha_{i})t/T\right) - (\Tilde{\mathcal{E}} - \mathcal{E}),
\end{equation}
where $\alpha_{i,f} = \text{tan}^{-1}[ (\Tilde{\mathcal{E}} - \mathcal{E} + \varepsilon(t=t_{i,f}))/2\sqrt{L}\mathcal{J}]$ must be close to $\pm\pi/2$ to approach the asymptote of the tangent. This function describes a rapid temporal evolution far from the central gap and a very slow evolution near it, ensuring that the adiabaticity condition is satisfied where the energy levels are closest. The geodesic path is specifically designed to minimize the energy variance during the system's evolution. By aligning the driving protocol with the metric tensor's structure, the fluctuations in energy are reduced, especially near the avoided crossing, where the gap is smallest. The probability for non-adiabatic transitions is minimized, insuring a smooth controlled evolution.

\subsection{Counterdiabatic evolution}
Choosing the temporal dependence of the protocol based on the system's geodesics enables optimal adiabatic driving along its eigenstates. However, adiabaticity inherently requires slow driving. To address this fundamental limitation and manipulate states as quickly as possible, a counterdiabatic Hamiltonian can be employed. The purpose of using a CD Hamiltonian is to exactly compensate for the terms responsible for non-adiabatic transitions when the system's temporal evolution violates the adiabaticity criterion \cite{Berry,Bason2012,Mandelstam,Chen2010,Martinez2014,Opatrný_2014}. In the same basis as in  Eq.(~\ref{gmunu}), namely the eigenbasis of the Bose-Hubbard Hamiltonian, the corrective term takes the form
\begin{equation}\label{hcd}
    H_{\text{CD}}(t) = i\hbar \sum_{m\neq n}\sum_{n} \frac{\langle m \vert\dot{H}(t)\vert n \rangle}{E_{n}-E_{m}}\vert m \rangle \langle n \vert.
\end{equation}
Thus, defining a new Hamiltonian $H + H_{\text{CD}}$ eliminates all off-diagonal terms in the eigenbasis, thereby preventing transitions to instantaneous eigenstates other than the one corresponding to the system's initial state. 

In general, it is challenging to compute the exact $H_{\text{CD}}$ for complex, large-scale systems. Several approximations exist to identify the most relevant terms for driving \cite{Sels,Claeys, approx1,approx2}. In our case, the mapping to a reduced system allows for the exact calculation of the $H_{\text{CD}}$ in the framework of the reduced Hamiltonian. For our system, the CD Hamiltonian takes the form
\begin{equation}\label{redhcd}
    \hat{H}_{\text{CD}}(t) =  \frac{\hbar \mathcal{N}\sqrt{L}\mathcal{J}\dot{\varepsilon}(t)}{4L\mathcal{J}^{2} + (\Tilde{\mathcal{E}}-\mathcal{E} + \mathcal{N}\varepsilon(t))^{2}} \hat{\sigma}_{y}
\end{equation} 
which we will denote as $\hat{H}_{\text{CD}}(t) = \hbar \Omega(t) \hat{\sigma}_{y}$ where $\hat{\sigma}_{y}$ is the Pauli matrix. The Hamiltonian $H_{\text{red}} + H_{\text{CD}}$ will thus dictate the system's temporal evolution, which will always satisfy the adiabaticity criterion by construction.

For an arbitrary driving function $\varepsilon(t)$, it is observed that the value of $H_{\text{CD}}$ depends on the time derivative of the driving protocol. Since the purpose of this corrective term is to compensate for the speed of evolution in cases where it might cause the system to exit the adiabatic regime, the elements of $H_{\text{CD}}$ must indeed take increasingly large values that may feature complicated time dependence. However, chosing a specific type of control allows to define the CD driving in terms of time-independent external fields \cite{electric}, which is advantageous from an experimental point of view since time-dependent protocols can make manipulation challenging and invariably require more resources. In particular, the usage of geodesic driving defines a counterdiabatic Hamiltonian whose elements are constant over time \cite{Dengis25}. Indeed, the essence of geodesic driving is to minimize its time derivative precisely around the central gap, where the $H_{\text{CD}}$ needs to take its largest values. Conversely, any counterdiabatic driving is of no practical use far away from the avoided crossing, where $\dot{\varepsilon}$ reaches its largest values. Thus, by substituting the expression for $\varepsilon(t)$ into Eq.(\ref{redhcd}), we obtain the GCD protocol 
\begin{equation}\label{omega}
    \Omega = \frac{\alpha_{f} - \alpha_{i}}{2T}.
\end{equation}
Remarkably, the elements of $H_{\text{CD}}$ are, as anticipated, independent of time (and also of the number of sites considered). The norm of $H_{\text{CD}}$ therefore depends solely on the total protocol time $T$, given that $\alpha_{f} - \alpha_{i} \approx -\pi$.

This result is justified by the form of $H_{\text{CD}}$. It is possible to demonstrate a relationship linking the counterdiabatic Hamiltonian and the metric tensor, namely through
\begin{equation}
\langle n | H_{\text{CD}}^{2} | n\rangle = -\hbar^{2}g^{(n)}\dot{\varepsilon}(t)^{2}.
\end{equation}
Thus, since $\hat{\sigma}_{y}^{2} = \hat{1}$, any $H_{\text{CD}}$ proportional to $\sigma_{y}$ will have time-independent elements if the driving $\varepsilon(t)$ satisfies the geodesic equation $g^{(n)}\dot{\varepsilon}^{2} = \text{const}$. Even more importantly, the geodesic equation is directly related to energy fluctuations along optimal trajectories. The Mandelstam-Tamm bound states that the standard time $\Delta t$ for energy to transition from one system state to another cannot be smaller than $\pi\hbar / 2\Delta E$ \cite{Mandelstam}, in the case of a final state that is orthogonal to the initial one \cite{deffner2017}. This fundamentally significant result provides a definition of the quantum speed limit, which is the maximum rate at which a system can evolve under an external perturbation. Applying this limit for our system, we find that the minimum time follows the inequality $T \ge \pi\hbar / 2\Omega$. This inequality imposes a constraint on the value of $\Omega$, which aligns precisely with the expression (\ref{omega}). We can thus conclude that the GCD protocol saturates the quantum speed limit.

The reduced Hamiltonian driving the system along its eigenstates in full adiabaticity is written as $H_{\text{red}} + H_{\text{CD}}$. The spectrum of this two-level Hamiltonian can be calculated exactly, and the energies are expressed as:
\begin{align}
    2E_{\pm} &= \Tilde{\mathcal{E}} + \mathcal{E} + \mathcal{N}\varepsilon(t) \\
    \notag &\pm \sqrt{4(L\mathcal{J}^{2} + \Omega^{2}) + (\Tilde{\mathcal{E}}-\mathcal{E} + \mathcal{N}\varepsilon(t))^{2}}.
\end{align}
Hence, the minimal gap will be expressed as $2\sqrt{L\mathcal{J}^{2} + \Omega^{2}}$. The presence of a term $\Omega^{2} \sim T^{-2}$ ensures that the gap is widened for short protocol times $T$, for which without the contribution of $H_{\text{CD}}$, the system would fall outside the adiabaticity criterion. Given that the time required for adiabaticity is inversely proportional to the gap, it is possible to estimate the time savings achieved with the use of the GCD protocol by comparing the minimal gap sizes:
\begin{equation}
    T_{G}/T_{GCD} \sim \left( 1+ \frac{\Omega^{2}}{L\mathcal{J}^{2}}\right)^{1/2}.
\end{equation}
With Eq.~(\ref{couplageeff}), we can therefore conclude that the proposed method must exhibit, compared to geodesic driving alone, a time-saving gain that behaves as $g = T_{G}/T_{GCD} \sim U^{N-1}/J^{N}$. The exponential dependence of the gain on the particle number in the proposed protocol is of crucial importance for any implementation that is of interest for quantum metrology, where $N$ should be as large as possible.
\subsection{Implementation in the many-body system}

To obtain an experimentally relevant protocol, it is necessary to emulate the action of the counterdiabatic Hamiltonian through the physical parameters of the system, namely the interaction $U$, the hopping $J$, and the driving $\varepsilon$. Various ways to emulate $H_\text{CD}$ were proposed \cite{SP,SP2,Campo2,petitziolpra}. Here, we define effective parameters $U_{\text{eff}}$, $J_{\text{eff}}$, and $\varepsilon_{\text{eff}}$ through the identification 
\begin{equation}
    H_{\text{red}}(U,J,\varepsilon(t)) + H_{\text{CD}}  \stackrel{!}{=}H_{\text{red}}\left(U_{\text{eff}}, J_{\text{eff}},\varepsilon_{\text{eff}}(t)\right). 
\end{equation}
The equations constraining the parameters can be fully determined via:
\begin{align}\label{jeff}
    &J_{\text{eff}} = U_{\text{eff}}^{1-1/N}\left[ J^{N}/U^{N-1} + i\hbar(N-1)!\Omega /N \right]^{1/N} \\
    \label{ueff}
    &U_{\text{eff}} = U + 2(\mathcal{E} - \mathcal{E}_{\text{eff}})/N(N-1)\\
    &\varepsilon_{\text{eff}}(t) = \mathcal{N}\varepsilon(t)/\mathcal{N}_{\text{eff}} + (\Tilde{\mathcal{E}} - \Tilde{\mathcal{E}}_{\text{eff}})/\mathcal{N}_{\text{eff}},
\end{align}
where $\Tilde{\mathcal{E}}_{\text{eff}}= \Tilde{\mathcal{E}}(U_\text{eff},J_{\text{eff}})$, $\mathcal{E}_{\text{eff}}= \mathcal{E}(U_\text{eff},J_{\text{eff}})$ and $\mathcal{N}_{\text{eff}} = \mathcal{N}(U_\text{eff},J_{\text{eff}})$ are the perturbative funcwtions evaluated with the effective parameters. This set of equations allows for the complete determination of the effective parameter values used to emulate the action of the counterdiabatic Hamiltonian. The strength of this result lies in the fact that, thanks to the GCD protocol, it becomes possible to accelerate the adiabatic creation of states using physical parameters that remain constant over time. Thus, in the self-trapping regime, the Bose-Hubbard Hamiltonian can be written as:
\begin{align}\label{BHMcd}
    \hat{H}(t)= &\frac{U_{\text{eff}}}{2}\sum_{i=1}^{L}\hat{a}_{i}^{\dagger}\hat{a}_{i}^{\dagger}\hat{a}_{i}\hat{a}_{i} 
    -\sum_{i=1}^{L}\left(J_{\text{eff }}\hat{a}^{\dagger}_{c}\hat{a}_{i} + J_{\text{eff }}^{*}\hat{a}^{\dagger}_{i}\hat{a}_{c}\right) \\
    \notag
    &+ \frac{U_{\text{eff}}}{2}\hat{a}_{c}^{\dagger}\hat{a}_{c}^{\dagger}\hat{a}_{c}\hat{a}_{c} +
     \varepsilon_{\text{eff}}(t)\hat{a}_{c}^{\dagger}\hat{a}_{c},
\end{align}
which will describe an adiabatic temporal evolution along the eigenstates of the Hamiltonian (\ref{BHM}), requiring only a modulation of the energy of the central well. Note that a complex hopping matrix element $J_{\text{eff}}\neq J_{\text{eff}}^{*}$ is required, which could be in practice implemented by a static gauge field creating  a phase $\phi = \text{tan}^{-1}\left( \Omega\hbar/\mathcal{J}\right)$ between sites via, for example, Floquet engineering \cite{Dalibard,Aidelsburger2011, Jimenez2012,Miyake13,Jotzu2014,Goldman2014,Sun2023,Petiziol2024} (see also the Supplemental Material of \cite{Dengis25}).

Globally, the overall structure of the spectrum is only slightly altered, as the action of the counterdiabatic term primarily focuses on widening the gap. However, there are limitations. Since the effective parameters depend on $\Omega$, which is itself inversely proportional to $T$, a protocol time that is too short may lead to complications. Indeed, proceeding too quickly would mean significantly altering the parameters and thus potentially deviating from the self-trapping regime. In particular, if the energy levels of interest are modified too strongly, they could become closer to the rest of the spectrum and thus exhibit interferences even with states lying a priori farther away in energy. The only contribution depending on the driving speed is $\Omega$, which must be compared to the initial separation between the considered levels and the states $\vert N-1;1,0,...\rangle, \vert N-1;0,1,...\rangle,...$, proportional to $U(N-1)/J$. Thus, knowing the expression of $\Omega$ in the case of geodesic counterdiabatic driving, we obtain a bound on the protocol time required to remain in the self-trapping regime: $ T\gg \pi J/2U(N-1)$. Therefore, although the quantum speed limit is reached for the reduced Hamiltonian, some fundamental limitations remain for the full Hamiltonian.
\section{Results and discussion}
\subsection{Triple-NOON state creation}

\begin{figure}[!t]
    \centering
    \includegraphics[width=0.475\textwidth]{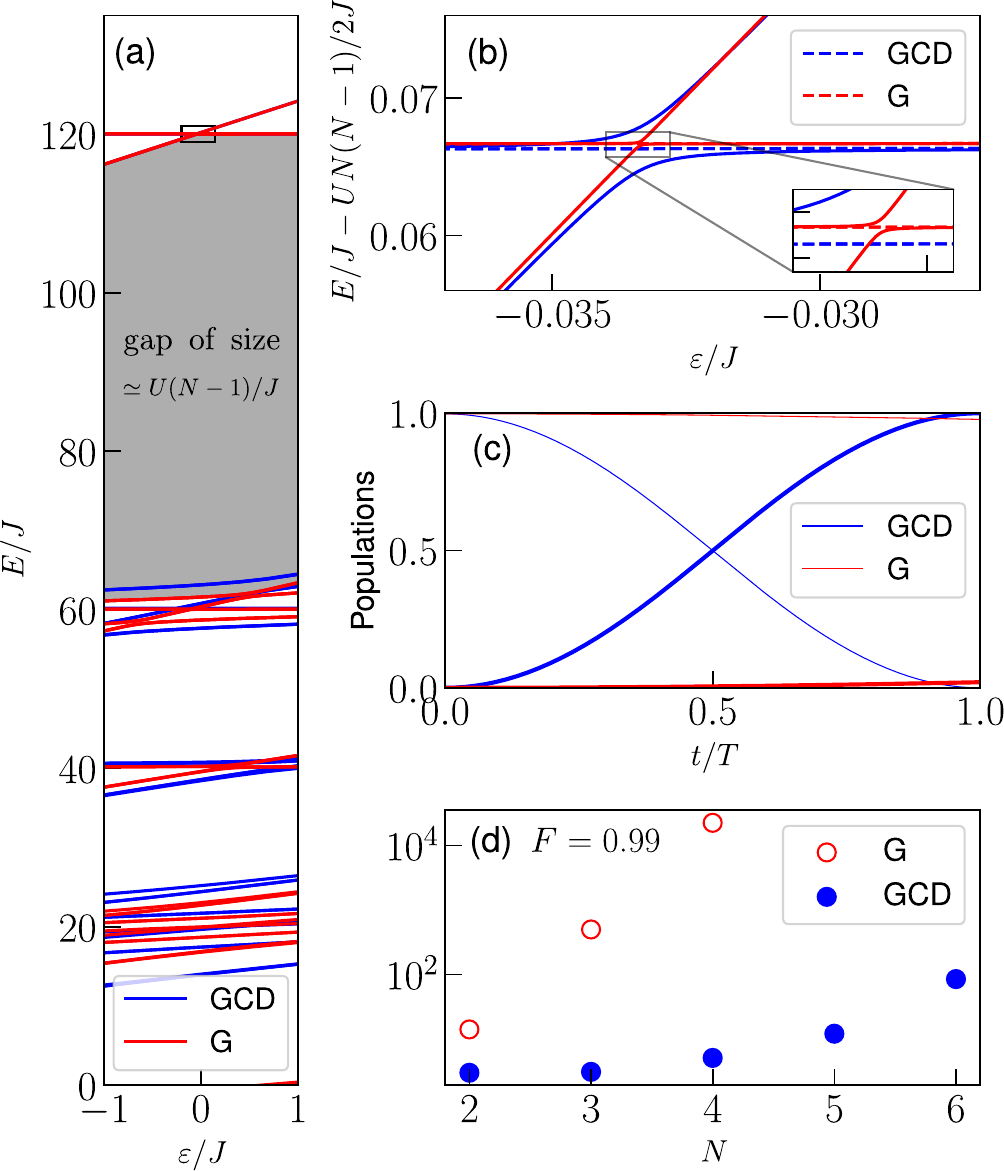}
    \caption{(a) Spectrum of the Bose-Hubbard Hamiltonian under geodesic driving (G, red curves) and geodesic counterdiabatic driving (GCD, blue curves) as a function of the parameter $\varepsilon /J$, for $U=20J$ and $N=4$. The states of interest, namely those where all particles are in the same level, are protected by a gap of value $U(N-1)$ from the rest of the spectrum, enabling the definition of a reduced system. A black rectangle highlights the avoided crossing region. (b) Zoom into the avoided crossing region. The gap is significantly widened due to the GCD protocol, allowing much faster adiabatic creation of triple-NOON states. (c) Population of the states $\vert N;0,0,0\rangle$ (thin lines) and $\vert \text{3-NOON} \rangle$ (thick lines) over time, for a total protocol time $T=10^{3}\hbar/J$. Population inversion is complete with the GCD protocol, whereas it does not even reach 10\% with simple geodesic driving (G). (d) Various values of the time required to create a triple-NOON state with G and GCD driving for different numbers of particles $N$, with a fixed $U=20J$ and a fixed fidelity $F=0.99$. An exponential growth in creation time is observed with the G protocol, while it is effectively mitigated with the GCD protocol, achieving experimentally feasible creation times.
    }
    \label{TNOON}
\end{figure}
 
To illustrate the effectiveness of the developed technique, we apply the protocol to the creation of highly entangled triple-NOON states, corresponding to the superposition $\vert \text{3-NOON}\rangle = (\vert N,0,0\rangle + \vert 0,N,0\rangle + \vert 0,0,N\rangle)/\sqrt{3}$. In our case, creating such a state involves considering a system with $L=3$ sites, where a central site is coupled to the three others. The energy of this central site will be modified as described by the GCD protocol presented earlier, to adiabatically generate the entangled state $\vert \text{3-NOON}\rangle$. Initially, the system is in the asymptotic Fock state $\vert N;0,0,0\rangle$, where all the bosons are located in the central well.

Figure \ref{TNOON}(a) qualitatively illustrates the effect of the GCD protocol on the spectrum of the Bose-Hubbard Hamiltonian. Specifically, a zoom (Fig.~\ref{TNOON}(b)) reveals the stretching of the avoided crossing, enabling much faster adiabatic transitions. The populations of the states $\vert N;0,0,0\rangle$ and $\vert 3-\text{NOON}\rangle$ over time are shown in Fig.~\ref{TNOON}(c) for $T=10^{3}\hbar/J$. A fidelity close to 100\% is achieved at the end of the GCD protocol, whereas only a small percentage of the population is found in the triple-NOON state with the G protocol. To provide a meaningful comparison for different particle numbers, Fig.~\ref{TNOON}(d) presents the time required to achieve a population of 0.99 in the triple-NOON state as a function of $N$. An exponential growth is observed for the G protocol, while it is mitigated with the GCD protocol, allowing for fast creation of triple-NOON states with ultracold atoms. As an example, such a state of $N = 4$ particles is realized with a fidelity of 0.99 in $\simeq 10^{2}\mathrm{s}$ using the G protocol for $U=20J$, whereas the GCD method requires only $0.02\mathrm{s}$. 

\subsection{Quantum Fisher information}
As previously mentioned, NOON states are particularly relevant for quantum metrology.
Here, we demonstrate the usefulness of our triple-NOON states through the calculation of the Quantum Fisher Information (QFI) associated with the system.
Let us consider the system's state in the presence of a phase shift between the wells,
\begin{align}
    \vert\psi(\theta_{1},\theta_{2},\theta_{3})\rangle =b \vert N;0,0,0\rangle + a ( e^{i\theta_{1}N} \vert 0;N,0,0\rangle \\
    \notag + e^{i\theta_{2}N} \vert 0;0,N,0\rangle
    + e^{i\theta_{3} N}\vert 0;0,0,N \rangle ). 
\end{align}
Here, we consider measuring the phase shift between each well relative to the central one, which evolves over time.
The normalization constants $\alpha$ and $\beta$ describe the population dynamics in the modes $\vert N;0,0,0 \rangle$ and $\vert \text{3-NOON}\rangle$, respectively, as illustrated in Fig.~\ref{TNOON}.
In the case of pure states, the Quantum Fisher Information (QFI) matrix is given by the expression

\begin{equation}
    F_{\mu\nu} = 4\text{Re}\left\{ \langle \partial_{\mu}\psi\vert \partial_{\nu}\psi\rangle -\langle \partial_{\mu}\psi\vert \psi\rangle \langle \psi \vert \partial_{\nu}\psi \rangle\right\},
\end{equation}
with $\partial_{\mu} \equiv \partial /\partial\theta_{\mu}$, where the indices $\mu$ and $\nu$ taking their values in $\{1, 2, 3\}$ correspond respectively to the phases $\theta_{1}$, $\theta_{2}$, and $\theta_{3}$.
The QFI is particularly useful for determining the lower bound on the variance of all phase parameters \cite{PRLNOON} :
\begin{equation}
    \vert \Delta\bm{\theta}\vert ^{2} = \sum_{m=1}^{3}\delta \theta_{m}^{2} \geq \text{Tr}[F_{\mu\nu}^{-1}].
\end{equation}
In particular, it was shown \cite{PRLNOON} that the minimum total phase variance is achieved for $a = 1/\sqrt{d + \sqrt{d}}$, and is given by $d(1+\sqrt{d})^{2}/(4N^{2})$, where $d$ denotes the number of estimated phases.
We refer to the NOON state corresponding to this optimal configuration as the Humphrey's NOON state.

\begin{figure}[t]
    \centering
    \includegraphics[width=0.475\textwidth]{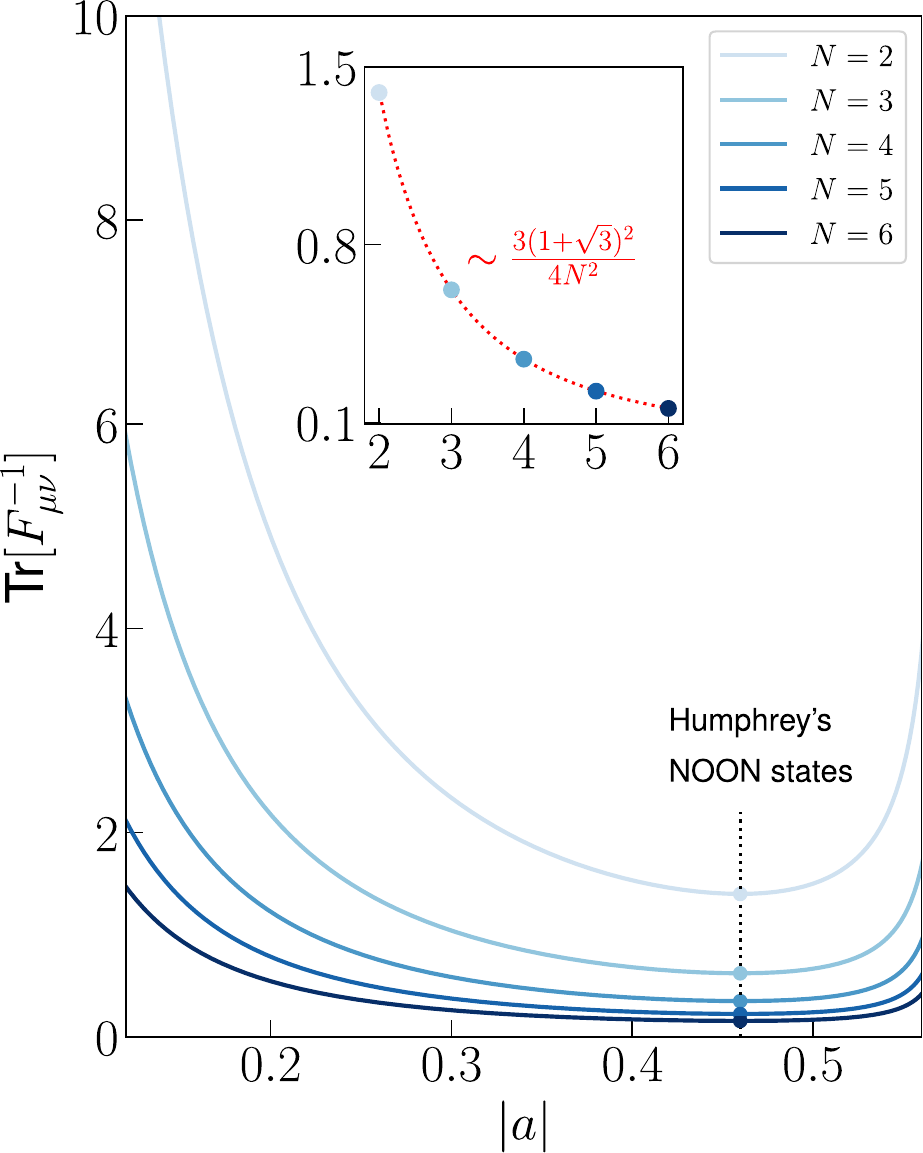}
    \caption{Evolution of the total phase variance for the state $\vert \psi(\theta_{1},\theta_{2},\theta_{3})\rangle$ as a function of the population root of the $\vert \text{3-NOON}\rangle$ state for different particle numbers $N$.
    This graph highlights the minimum of $\vert \Delta \bm{\theta}\vert ^{2}$ predicted by Humphrey {\em et al.} \cite{PRLNOON} at $\alpha = 1/\sqrt{d + \sqrt{d}}$.
    In the inset, minimal values of $\text{Tr}[F_{\mu\nu}^{-1}]$ are plotted as a function of the number of particles to confirms the Heisenberg scaling of $\sim 1/N^{2}$ for pure NOON states, which occurs when the GCD protocol is fully implemented} 
    \label{Fish}
\end{figure}

In Fig.~\ref{Fish}, we examine the evolution of the QFI as a function of $\vert a \vert$, for different particle numbers, in the context of generating a $\vert \text{3-NOON}\rangle$ state.
In particular, we demonstrate that our state preparation method passes through the optimal point predicted by Humphrey {\em et al.} \cite{PRLNOON}, and that the final value of $\vert \Delta \bm{\theta}\vert ^{2}$ reaches the Heisenberg limit with particle number, namely $3(1+\sqrt{3})^{2}/4N^{2}$.

\subsection{L-NOON states creation}

\begin{figure}[!t]
    \centering
    \includegraphics[width=0.475\textwidth]{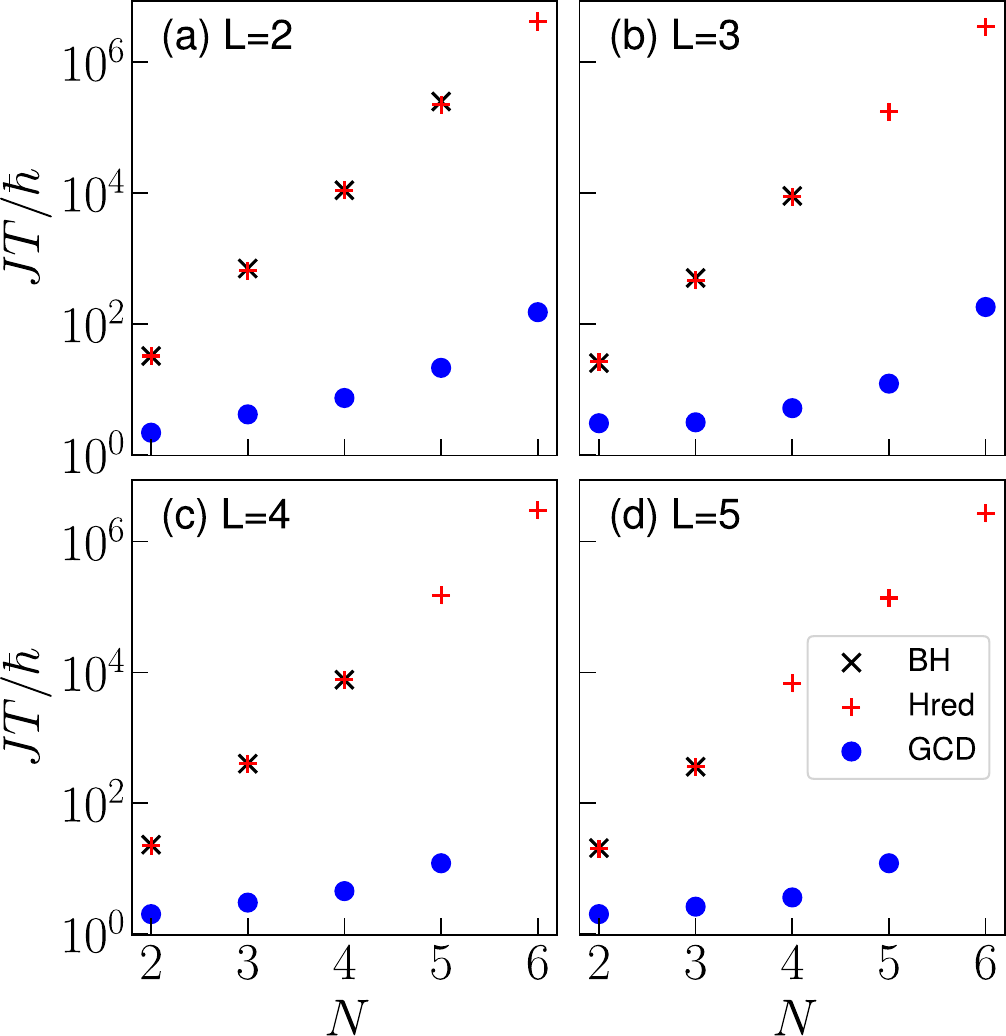}
    \caption{Protocol time $JT/\hbar$ required to achieve 99\% fidelity of the state (a) NOON, (b) 3-NOON, (c) 4-NOON, and (d) 5-NOON, in logarithmic scale, for a geodesic driving (Bose-Hubbard system in black (x), reduced system in red (+), and for the geodesic counterdiabatic driving (GCD protocol in blue), with a fixed value of $NU/J=60$. Clearly, the GCD protocol allows for a gain of several orders of magnitude in the time required to create an $L$-NOON state. In particular, the exponential dependence of $T$ on the number of particles is very clearly visible. 
    }
    \label{LNOON}
\end{figure}

The GCD protocol being perfectly defined for any number of wells $L$, the extension of the results to more exotic states such as 4-NOON or 5-NOON is almost immediate. Figure \ref{LNOON} shows, for different values of $L$ and $N$, the time required to achieve 99\% fidelity with the desired state. The GCD protocol unequivocally allows the creation of large entangled states, such as the state $\vert \text{5-NOON}\rangle = (\vert 0;N,0,0,0,0\rangle + \vert 0;0,N,0,0,0\rangle + ... + \vert 0;0,0,0,0,N\rangle)/\sqrt{5}$, much faster than with a geodesic driving without counterdiabatic assistance. The reduced model easily allows for an estimate of the parameters and the protocol time, even for Bose-Hubbard systems of large size that are beyond computational capabilities.

As can be observed in Fig. \ref{LNOON}, the time required to obtain a $\vert \text{L-NOON}\rangle$ state appears to depend only weakly on the number of wells. This can be readily understood by noting from Hamiltonian (\ref{Hred}) that the effective coupling, which is directly proportional to the inverse time required for adiabaticity, is only modified by a factor of $\sqrt{L}$ as compared to its value for $L=1$. Given that this coupling decreases exponentially with the number of particles, such a factor is not significant as compared to the strong $N$ dependence.

It appears that the time needed to create the NOON state depends only very slightly on the number of wells, and in particular, this time decreases with $L$, coming from the fact that the effective coupling scales with $L^{1/2}$. Consequently, creating large entangled states is possible in even shorter time.

\subsection{Experimental challenges}

\begin{figure}[!t]
    \centering
    \includegraphics[width=0.475\textwidth]{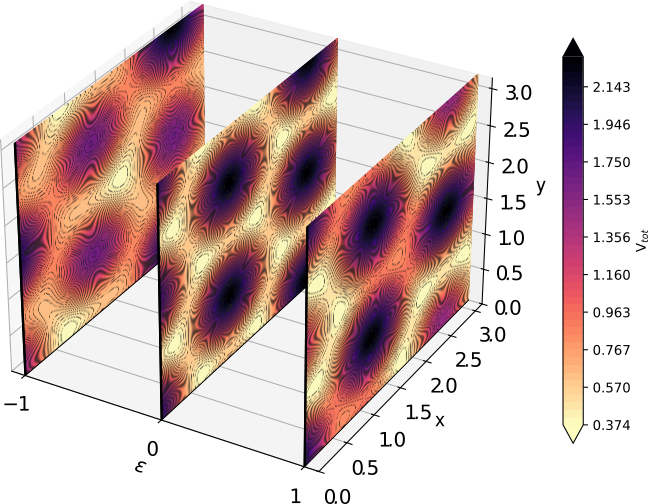}
    \caption{Schematic representation of a hexagonal optical lattice, in which the proposed protocol could be used to generate triple-NOON states. A triangular lattice is superposed on the latter in order to have control on the depth of the central well.  Initially, the bosons are trapped in one of the highest-energy wells ($\varepsilon=1$), effectively isolating the particles from other energy levels. The protocol then provides a method for modifying the optical lattice to control the energy of the central well using an additional external field. }
    \label{optical}
\end{figure}

Although the theoretical framework presents no fundamental obstacle to extending the protocol to larger numbers of particles or wells, certain experimental constraints must be taken into account. First, perturbation theory is inherently limited by the chosen maximum order. The larger the number of bosons involved in the system, the greater the precision required in determining the energy splitting to ensure adiabatic evolution. Additionally, it is necessary to maintain an almost perfect symmetry among the $L$ outer wells, a condition that becomes increasingly difficult to satisfy as $N$ increases.

The method proposed here is readily extensible numerically, as the only required parameters can be obtained through straightforward diagonalization of the Bose-Hubbard Hamiltonian. The precision on the energy splitting can therefore reach machine-level accuracy for any number $L$ of outer wells. However, the experimental realization of such a system may become increasingly challenging as $L$ grows. Several well-known lattice geometries can be employed to generate exotic states (a triangular lattice for $L=3$ \cite{Windpassinger2013}, a square for $L=4$ \cite{Bloch2008}, or a hexagonal lattice for $L=6$ \cite{Jotzu2014}).

\begin{figure}[!t]
    \centering
    \includegraphics[width=0.475\textwidth]{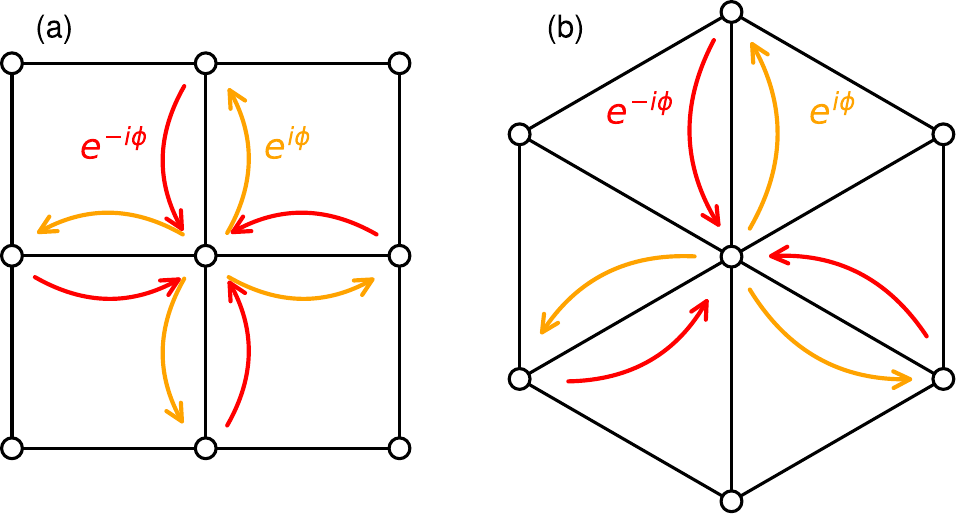}
    \caption{Examples of different geometries that could be used to realize the proposed method within (a) square and (b) trigonal lattices. In order to implement the complex hopping defined by Eq.~\ref{jeff}, it is necessary that a phase $\pm\phi$ exists between the sites. Specifically, all hoppings from the central site to the outer wells must carry the same phase sign, and the opposite phase sign applies for hoppings from the outer wells back to the central site. }
    \label{lattices}
\end{figure}

Another challenge, particularly within optical lattice setups, is to realize, via Floquet engineering \cite{Dalibard,Aidelsburger2011,Jimenez2012,Miyake13,Jotzu2014,Goldman2014,Sun2023,Petiziol2024} or related techniques, a complex hopping in the effective parameters of the Bose–Hubbard model. In our protocol, the same phase is required from the central site to all other sites, which does not correspond to a simple Peierls phase with alternating signs. Figure \ref{lattices} illustrates two possible cases for state creation: (a) a 4-NOON state in a square lattice and (b) a 3-NOON state in a triangular lattice, together with the associated phases required to emulate the action of the counterdiabatic Hamiltonian. Our GCD approach mitigates the experimental difficulty, since it relies on time-independent parameters. 

The larger $L$ becomes, the more problematic it is to neglect couplings between the outer wells. Still, one can quickly see that the presence of such couplings introduces only minor complications. Indeed, if an effective coupling $\mathcal{J'}$ between the $L$ outer wells is considered, the system can still be reduced to an effective two-level description. The energy of the outer wells will be shifted by a factor $-(L-1)\mathcal{J'}$, leading to a slight modification of Eq.~(\ref{ueff}) in the form of a constant shift that has no impact on the determination of the effective parameters. Since all outer wells are equally coupled to one another, this coupling acts as an internal energy for the $\vert \text{L-NOON}\rangle$ state.

The reduction to an effective two-level system is, however, only possible if the outer wells are perfectly symmetric with respect to one another. This symmetry requirement is difficult to achieve in traditional static optical lattices (see Appendix B for a quantitative discussion on the effect of disorder in the configuration). However, it can be achieved in the context of discrete time crystals, where such symmetry is guaranteed by construction \cite{wilczek2012,Sacha2015,Guanghui25}. In this case, the modes are located on the resonance islands of phase space \cite{Tomsovic1994,Brodier2001,peter1}. The method to be applied is adiabatic transfer from a central state symmetrically coupled to all other resonant modes which is localized at the center of the main regular island in phase space, the goal is thus to construct a nonlinear resonance centered around the central well, enabling symmetric tunneling to all other modes. This method could potentially be implemented in single optical lattice wells subjected to periodic modulation.

\section{Conclusion}
We presented an innovative method for the rapid and high-fidelity generation of multi-mode NOON states using ultracold atoms in lattice structures in the self-trapping regime. By combining an optimized geodesic driving protocol with a counterdiabatic Hamiltonian, we demonstrated that the quantum speed limit can be achieved, enabling efficient preparation of entangled states. Our approach relies on a rigorous reduction of the system and the emulation of effective time-independent parameters to adapt the protocols to experimental constraints. The application of this method to the creation of exotic 3,4,5-NOON states highlighted significant gains in both time and fidelity, paving the way for experimentally feasible implementations in complex large-scale systems.

\begin{acknowledgments}
This project (EOS 40007526) has received funding from the FWO and F.R.S.-FNRS under the Excellence of Science (EOS) programme. S.W. acknowledges support by Q-DYNAMO
(EU HORIZON-MSCA-2022-SE-01) with project No.
101131418 and by the National Recovery and Resilience
Plan "PNRR" through i) PRIN 2022 project "Quantum Atomic Mixtures: Droplets, Topological Structures, and Vortices", project No. 20227JNCWW, CUP D53D23002700006, and ii) Mission 4 Component 2 Investment 1.3
– Call for tender No. 341 of 15/03/2022 of Italian
MUR funded by NextGenerationEU, with project No.
PE0000023, Concession Decree No. 1564 of 11/10/2022
adopted by MUR, CUP D93C22000940001, Project title
“National Quantum Science and Technology Institute“
(NQSTI).
\end{acknowledgments}

\medskip
\appendix
\section{More on perturbative developpement}

In this section, we further develop the perturbative expansion in order to gain insight into the quality of the approximation that we employed. The calculations will be restricted to the case $L=2$, for which a quantitative estimate of the possible errors can be obtained.

\subsection{External wells}
The equations to be solved in order to obtain the sixth-order perturbative corrections to the energies of the outer wells remain relatively manageable, as the number of equations involved is still accessible. This time, we consider a six-step virtual process, corresponding for instance to a particle that moves three times and then returns. Let us denote the interaction energy shifted by the driving term $\varepsilon$ as
\begin{equation}\label{EN}
    E_{n_{0};n_{1},n_{2}} = n_{0}\varepsilon + \sum_{i=0}^{2} U n_{i}(n_{i}-1)/2,
\end{equation}
and the hopping terms as 
\begin{equation}\label{JN}
\mathcal{J}_{n_{i}, n_{j}} = J\sqrt{n_{i}(n_{j}+1)},
\end{equation}
where $n_{i}$ are the number of particles at site $i$. We have
\begin{widetext}
\begin{align}
    &(E-E_{0;N,0})\Psi_{0;N,0} = -\mathcal{J}_{N,0}\Psi_{1;N-1,0}\\
    &(E^{(4)}-E_{1;N\!-\!1,0})\Psi_{1;N\!-\!1,0} = -\mathcal{J}_{1,N-1}\Psi_{0;N,0}-\mathcal{J}_{N-1,1}\Psi_{2;N\!-\!2,0} - \mathcal{J}_{1,0}\Psi_{0;N\!-\!1,1}\\
    &(E^{(2)}-E_{0;N\!-\!1,1})\Psi_{0;N\!-\!1,1} = -\mathcal{J}_{1,0}\Psi_{1;N\!-\!1,0} -\mathcal{J}_{N-1,0}\Psi_{1;N\!-2,1} \\
    &(E^{(2)}-E_{2;N\!-\!2,0})\Psi_{2;N\!-\!2,0} = -\mathcal{J}_{2,N\!-\!2}\Psi_{1;N\!-\!1,0} -\mathcal{J}_{2,0}\Psi_{1;N\!-\!2,1} -\mathcal{J}_{N\!-\!2,2}\Psi_{3;N\!-\!3,0}\\
    &(E^{(0)}-E_{1;N\!-\!2,1})\Psi_{1;N\!-\!2,1} = -\mathcal{J}_{1,1}\Psi_{2;N\!-\!2,0}-\mathcal{J}_{1,N\!-\!2}\Psi_{0;N\!-\!1,1} \\
    &(E^{(0)}-E_{3;N\!-\!3,0})\Psi_{3;N\!-\!3,0} = -\mathcal{J}_{3,N\!-\!3}\Psi_{2;N\!-\!2,0}
\end{align}
\end{widetext}
where $E^{(4)}$, $E^{(2)}$ and $E^{(0)}=UN(N-1)/2$ are the energies of $\vert0;N,0\rangle$ at fourth, second and zeroth order, respectively. Once solved and expanded in the regime $NU \gg \varepsilon, J$, we obtain
\begin{widetext}
\begin{align}
    \mathcal{E}^{(6)}=&\frac{NJ^2}{U(N-1)} +\frac{NJ^2\varepsilon}{U^2(N-1)^2} + \frac{NJ^4}{(N-1)^2(N-2)U^3} + \frac{NJ^4\varepsilon(3N^2 -7N + 3)}{(N-1)^4(N-2)^2U^4}
    +\frac{NJ^6(2N^3 - 6N^2 + N +5)}{(N-1)^5(N-2)^2(N-3)U^5} \\
    \notag &- \frac{NJ^6\varepsilon(16N^8 - 250N^7 + 1570N^6 -515N^5 + 9425N^4 -9155N^3 -3351N^2 + 1104N-918)}{(2N^2-9N-9)^2(N-1)^6(N-2)^3U^6}
\end{align}
\end{widetext}
The extension to complex $J$ is straightforward, replacing $J^2$ by $\vert J\vert ^2$.
\subsection{Central well}

\begin{figure}
    \centering
    \includegraphics[width=0.475\textwidth]{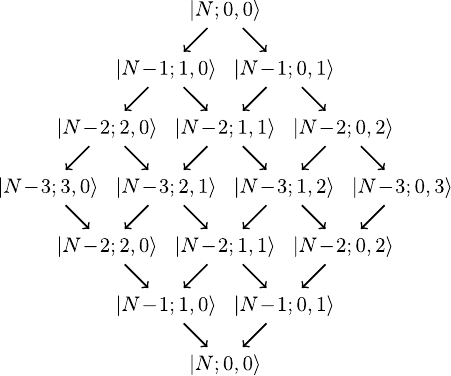}
    \caption{Perturbative diagram for the central well at sixth order for $L=2$. Each arrow corresponds to a probabilistic weight that characterizes the path taken by the system’s state across the different states of the Hilbert space.}
    \label{diago6}
\end{figure}

The sixth-order perturbative diagram displayed in Fig. \ref{diago6} illustrates the increasing complexity of the calculations. Since the number of trajectories is significantly larger, the corresponding equations to be solved are also more numerous.

However, it is possible to fold the diagram onto itself by defining the states according to the symmetry of the system. Specifically, by setting $\vert 1\rangle \equiv \vert N\! -\! 1;1,0\rangle + \vert N\!-\!1;0,1\rangle$, $\vert 2\rangle \equiv \vert N\!-\! 2;2,0\rangle + \vert N!\-\!2;0,2\rangle$, $\vert 3\rangle \equiv \vert N\!-\! 3;3,0\rangle + \vert N\!-\!3;0,3\rangle$, and $\vert 321\rangle \equiv \vert N\!-\! 3;2,1\rangle + \vert N\!-\!3;1,2\rangle$, the number of equations can be significantly reduced. Expanding for a bias $\varepsilon$ and in the self-trapping regime (i.e., $NU \gg \varepsilon,J$), we obtain

\begin{figure*}[!t]
    \centering
    \includegraphics[width=0.9\textwidth]{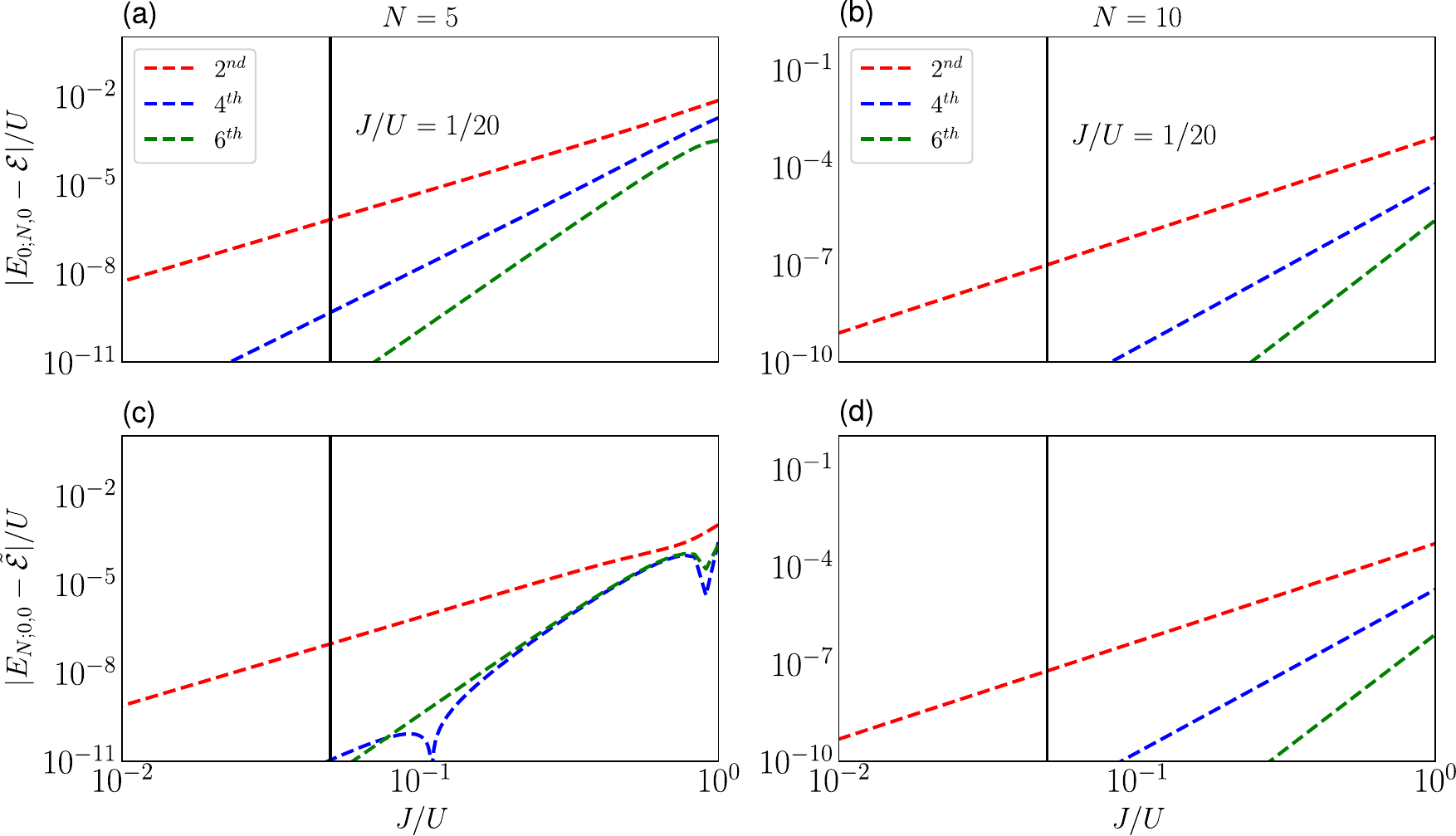}
    \caption{Convergence of the perturbative expansions for the two highest energies in the Bose–Hubbard model \ref{BHM} as a function of $J/U$. Panels (a),(b) display the relative deviation for the external wells $|E_{0;N,0} - \mathcal{E}|$ for $N=5$ and $N=10$, respectively, while panels (c),(d) show the deviation for the central well $|E_{N;0,0} - \tilde{\mathcal{E}}|$. The red, blue, and green dashed lines correspond to the $2^{\text{nd}}$, $4^{\text{th}}$, and $6^{\text{th}}$ order expansions. The black vertical line indicates the regime $J/U = 1/20$, which is used in our simulations. }
    \label{perturbativepic}
\end{figure*}

\begin{widetext}
\begin{align}
    &\Tilde{\mathcal{E}}^{(6)} = \frac{2NJ^2}{U(N-1)} -\frac{2NJ^2 \varepsilon}{U^2(N-1)^2} - \frac{2NJ^4(N^2 - 6N +7)}{(N-1)^3(N-2)(2N-3)U^3} -  \frac{2NJ^4\varepsilon(25N^3-123N^2+201N-109)}{(N-1)^4(2N^2-7N+6)^2U^4}  \\
    \notag &-\frac{2NJ^6 (12N^6 - 124N^5 + 381N^4 - 80N^3-1602N^2 + 2804N-1455)}{(2N-3)^2(N-3)(N-2)^2(N-1)^5(3N-7)U^5} \\
    \notag &+\frac{2NJ^6\varepsilon}{(3N-7)^2(N-3)^2(N-2)^3(N-1)^6(2N-3)^3 U^6}(48N^{12} - 1248N^{11}+14978N^{10} - 110137N^9+548649N^8 \\
   \notag &-1923443N^7 + 4781377N^6) -8313943N^5 - 9733925N^4 - 7041789N^3 + 2460637N^2 + 96384N - 247230).
\end{align}
\end{widetext}
The truncation error for the fourth order of perturbation theory is therefore determined up to the next order. Figure (\ref{perturbativepic}) shows the modulus of the differences between the well energies and their perturbative approximations at different orders, for $L=2$, as a function of the ratio $J/U$. Panels (a) and (b) compare the energy of an outer well for $N=5$ and $N=10$, respectively, while panels (c) and (d) correspond to the central well. One can see from Figure (\ref{perturbativepic}) that for the fourth order of perturbation theory, precision is already better than $10^{-8}$. 

\medskip
\section{Disorder and different couplings}
Here, we also study the impact of a well-depth fluctuation as well as of different couplings between the wells. We therefore consider a coupling $J$ between the central well and the left well, and a coupling $G$ with the right well. The energies are then written as $E_{n_{0};n_{1},n_{2}} = \sum_{i=0}^{2} n_i (\varepsilon_i + U(n_{i}-1)/2\big)$, where the $\varepsilon_i$ are the on-site energies of the different wells. The equations for the fourth-order corrections of the central well then become
\begin{widetext}
\begin{align}
    &(E-E_{N;0,0})\Psi_{N;0,0} = -\mathcal{J}_{N,0}\Psi_{N\!-\!1;1,0} -\mathcal{G}_{N,0}\Psi_{N\!-\!1;0,1} \\
    &(E^{(2)}-E_{N\!-\! 1;1,0})\Psi_{N\!-\!1;1,0} = -\mathcal{J}_{1;N\!-\!1}\Psi_{N;0,0}-\mathcal{J}_{N\!-\!1,1}\Psi_{N\!-\!2;2,0}-\mathcal{G}_{N\!-\!1,0}\Psi_{N\!-\!2;1,1} \\
    &(E^{(2)}-E_{N\!-\!1;0,1})\Psi_{N\!-\!1;0,1}=-\mathcal{G}_{1,N\!-\!1}\Psi_{N;0,0} - \mathcal{G}_{N\!-\!1,1}\Psi_{N\!-\!2,0,2}-\mathcal{J}_{N\!-\!1,0}\Psi_{N\!-\!2,1,1} \\
    &(E^{(0)}- E_{N\!-\!2;2,0})\Psi_{N\!-\!2,2,0} = -\mathcal{J}_{2,N\!-\!2}\Psi_{N\!-\!1;1,0} \\
    &(E^{(0)}-E_{N\!-\!2;0,2})\Psi_{N\!-\!2;0,2}=-\mathcal{G}_{2,N\!-\!2}\Psi_{N\!-\!1;0,1} \\
    &(E^{(0)}-E_{N\!-\!2;1,1})\Psi_{N\!-\!2;1,1}=-\mathcal{J}_{1,N\!-\!2}\Psi_{N\!-\!1;1,0} - \mathcal{G}_{1,N\!-\!2}\Psi_{N\!-\!1;0,1}
\end{align}
\end{widetext}
where $\mathcal{J}_{n_{i}, n_{j}}$ is defined by Eq.\ref{JN}, and $\mathcal{G}_{n_{i}, n_{j}} = G\sqrt{n_{i}(n_{j}+1)}$ is the probabilistic weight associated with the motion of a particle between the central well and the right well. Solving these equations leads to the following correction:
\begin{widetext}
\begin{align}\label{dis6c}
    \notag&\Tilde{\mathcal{E}}^{(4)}= \frac{\mathcal{J}_{N,0}\mathcal{J}_{1,N\!-\!1}}{E^{(0)}-E_{N\!-\!1;1,0}} - \frac{\mathcal{J}_{N,0}\mathcal{J}_{1,N\!-\!1}}{(E^{(0)}-E_{N\!-\!1;1,0})^2}\left(\frac{\mathcal{J}_{N,0}\mathcal{J}_{1,N\!-\!1}}{E^{(0)}-E_{N\!-\!1;1,0}} + \frac{\mathcal{G}_{N,0}\mathcal{G}_{1,N\!-\!1}}{E^{(0)}-E_{N\!-\!1;0,1}} - \frac{\mathcal{J}_{N\!-\!1,1}\mathcal{J}_{2,N\!-\!2}}{E^{(0)}-E_{N\!-\!2;2,0}} - \frac{\mathcal{G}_{N\!-\!1,1}\mathcal{G}_{1,N\!-\!2}}{E^{(0)}-E_{N\!-\!2;1,1}} \right) \\
     \notag&+\frac{\mathcal{G}_{N,0}\mathcal{G}_{1,N\!-\!1}}{E^{(0)}-E_{N\!-\!1;0,1}} - \frac{\mathcal{G}_{N,0}\mathcal{G}_{1,N\!-\!1}}{(E^{(0)}-E_{N\!-\!1;0,1})^2}\left(\frac{\mathcal{J}_{N,0}\mathcal{J}_{1,N\!-\!1}}{E^{(0)}-E_{N\!-\!1;1,0}} + \frac{\mathcal{G}_{N,0}\mathcal{G}_{1,N\!-\!1}}{E^{(0)}-E_{N\!-\!1;0,1}} - \frac{\mathcal{G}_{N\!-\!1,1}\mathcal{G}_{2,N\!-\!2}}{E^{(0)}-E_{N\!-\!2;0,2}} - \frac{\mathcal{J}_{N\!-\!1,0}\mathcal{J}_{1,N\!-\!2}}{E^{(0)}-E_{N\!-\!2;1,1}} \right) \\
     &+\frac{\mathcal{J}_{N,0}\mathcal{G}_{N\!-\!1,0}\mathcal{G}_{1,N\!-\!2}\mathcal{J}_{1,N\!-\!1}}{(E^{(0)}-E_{N\!-\!1;1,0})(E^{(0)}-E_{N\!-\!1;0,1})(E^{(0)}-E_{N\!-\!2;1,1})} +\frac{\mathcal{G}_{N,0}\mathcal{J}_{N\!-\!1,0}\mathcal{J}_{1,N\!-\!2}\mathcal{G}_{1,N\!-\!1}}{(E^{(0)}-E_{N\!-\!1;1,0})(E^{(0)}-E_{N\!-\!1;0,1})(E^{(0)}-E_{N\!-\!2;1,1})}.
\end{align}
\end{widetext}

In this form, it is clear that each term corresponds to a different trajectory in Hilbert space. For instance, the term proportional to $\mathcal{J}_{N,0}^2 \mathcal{J}_{1,N\!-\!1}^2$ corresponds to a particle starting from the central well and performing two round trips to the left well. 
For the outer wells, the expression is much simpler due to the presence of only one single loop in the perturbative diagram, as opposed to three in the case of the central well. We have
\begin{align}\label{dis6e}
    &\mathcal{E}^{(4)} = \frac{\mathcal{J}_{N,0}\mathcal{J}_{1,N\!-\!1}}{E^{(0)}-E_{1;N\!-\!1,0}}-\frac{\mathcal{J}_{N,0}\mathcal{J}_{1,N\!-\!1}}{(E^{(0)}-E_{1;N\!-\!1,0})^2} \\
    \notag &\times\left(\frac{\mathcal{J}_{N,0}\mathcal{J}_{1,N\!-\!1}}{E^{(0)}-E_{1;N\!-\!1,0}} - \frac{\mathcal{J}_{N\!-\!1,1}\mathcal{J}_{2,N\!-\!2}}{E^{(0)}-E_{2;N\!-\!2,0}} -\frac{\mathcal{G}_{1,0}\mathcal{G}_{1,0}}{E^{(0)}-E_{0;N\!-\!1,1}} \right).
\end{align}
From Eqs. \ref{dis6c} and \ref{dis6e}, it is possible to include the energy differences between the central well and the outer wells, thereby capturing the notion of disorder in the optical lattice. To picture the behaviour of disorder, we consider small deviations on the hopping ($G\rightarrow J+\delta J/2$, $J \rightarrow J -\delta J/2$) and on onsite energies of external wells ($E_{0;N,0} - E_{0;0,N} = N(\delta\epsilon_1 - \delta\epsilon_2)$). We can then develop expressions \ref{dis6c} and \ref{dis6e} to obtain, to the second order of disorder, the energies for the central well ($\Tilde{\mathcal{E}^{(2)}}$), for the left ($\mathcal{E}^{(2)}_1$) and right ($\mathcal{E}^{(2)}_2$) wells:

\begin{widetext}
\begin{align}
\Tilde{\mathcal{E}}^{(2)} &\approx \frac{2NJ^2}{U(N-1)}+ \frac{NJ^2}{(N-1)^2U^2}(\delta\epsilon_1 + \delta\epsilon_2) + \frac{NJ^2}{(N-1)^3 U^3}(\delta\epsilon_1^2+\delta\epsilon_2^2) -\frac{NJ}{(N-1)^2U^2}(\delta\epsilon_1 +\delta \epsilon_2)\delta J+\frac{N}{2U(N-1)}\delta J^2 \\
\mathcal{E}^{(2)}_1 &\approx \frac{NJ^2}{U(N-1)}-\frac{NJ^2}{(N-1)^2U^2}\delta\epsilon_1 + \frac{NJ^2}{(N-1)^3U^3}\delta\epsilon_1^2 -\frac{NJ}{(N-1)U}\delta J + \frac{NJ}{(N-1)^2U^2}\delta\epsilon_1\delta J + \frac{N}{4U(N-1)}\delta J^2 \\
\mathcal{E}^{(2)}_2 &\approx \frac{NJ^2}{U(N-1)}-\frac{NJ^2}{(N-1)^2U^2}\delta\epsilon_2 + \frac{NJ^2}{(N-1)^3U^3}\delta\epsilon_2^2 +\frac{NJ}{(N-1)U}\delta J - \frac{NJ}{(N-1)^2U^2}\delta\epsilon_2\delta J + \frac{N}{4U(N-1)}\delta J^2
\end{align}
\end{widetext}
To give a concrete example, we consider a symmetrical detuning around the energy of the central well, that has the form $\delta\epsilon_1 = -\delta\epsilon_2 = \delta\epsilon/2$ :
\begin{widetext}
\begin{align}
    \Tilde{\mathcal{E}}^{(2)} &\approx \frac{2NJ^2}{U(N-1)}+ \frac{NJ^2}{2(N-1)^3 U^3}\delta\epsilon^2 + \frac{N}{2(N-1)U}\delta J^2 - \frac{NJ}{(N-1)^2 U^2}\delta J \delta \epsilon, \\
    \mathcal{E}^{(2)}_1 &\approx \frac{NJ^2}{U(N-1)}-\frac{NJ^2}{2(N-1)^2 U^2}\delta\epsilon + \frac{NJ^2}{4(N-1)^3U^3}\delta\epsilon^2-\frac{NJ}{U(N-1)}\delta J + \frac{NJ}{2(N-1)^2 U^2}\delta J \delta \epsilon + \frac{N}{(N-1)^2 U^2}\delta J^2, \\
    \mathcal{E}_{2}^{(2)} &\approx \frac{NJ^2}{U(N-1)}+\frac{NJ^2}{2(N-1)^2 U^2}\delta\epsilon + \frac{NJ^2}{4(N-1)^3U^3}\delta\epsilon^2+\frac{NJ}{U(N-1)}\delta J + \frac{NJ}{2(N-1)^2 U^2}\delta J \delta \epsilon + \frac{N}{(N-1)^2 U^2}\delta J^2.
\end{align}
\end{widetext}
This yields a quantitative information on the extent to which the presence of disorder can be tolerated without affecting the effectiveness of the protocol.

%\bibliography{biblio.bib}

\providecommand{\noopsort}[1]{}\providecommand{\singleletter}[1]{#1}%

\end{document}